# Time-resolved Diffusing Wave Spectroscopy to Enable Ultrafast Electro-optic Relaxation Dynamic in Liquid Crystals

Suman Kalyan Manna[1*], Laurent Dupont[2], Guoqiang Li[3*]

[1]Department of Cell Biology and Human Anatomy, University of California, Davis, 95616, USA
[2]Department of Optics, Telecom Bretagne, 655 Avenue du Technopole, 29200 Plouzané, France
[3]Department of Electrical and Computer Engineering, The Ohio State University, Columbus, OH 43212, USA

Enabling control over a spontaneous dynamic process is a very complicated task in practice because of its statistical nature, and electro-optic (E-O) relaxation dynamic in a nematic liquid crystal (NLC) system is not an exception. Controlling E-O relaxation time requires an appropriate microscopic visualization. For that, a time-resolved diffusing wave spectroscopy (TR-DWS) is developed first time, and further demonstrated with that the linear E-O relaxation time of a NLC can be improved typically, from 25msec to 350$\mu sec$ at room temperature, by controlling the orientational ordering of NLC. Apart from improving the linear E-O dynamic, it is expected that the application range of the proposed TR-DWS is likely to be extended to applications such as optical imaging in turbid media, coherent controlling and optical manipulation in complex quantum systems including miniature lasers, amorphous photonic crystals, and more.

- ## INTRODUCTION

The E-O relaxation dynamic of a NLC system is considered to be a statistically random dynamic process [1-3]. During the transition of an externally applied electric field from ON to OFF, LC molecules relax back from homeotropic to planar state in presence of a non-uniform surface energy having ideally, a maximum value ($w \to \infty$) at the surface ($Z = 0$) of the cell, and a minimum value $w \to 0$ at the middle ($Z = d/2$) of an infinitely thick planar cell[2,3]. This non-uniformity in surface energy leads to a distribution in polar angle $\vartheta = \vartheta_m sin(d/\xi)$ (**Fig.1a**) of the local directors of LC molecules, where $\vartheta_m$ is the maximum angle of a local director, situated at the middle of the cell, $d$ is the thickness of the cell, and $\xi = \frac{d}{\pi}$ is the correlation length of the local director orientations[4,5]. The time evolution of $\vartheta$ during the relaxation process gives rise to the fluctuation in optical phase of an optical beam passing through the cell [3-5]. As a result, the relaxation process of a NLC is found to be a statistically random dynamic process [3,5]. It might be interesting to analyze such a random temporal-dynamic process, manifested during the E-O relaxation in terms of diffusing wave spectroscopy (DWS), a technique usually applied to some turbid media to calculate some physical parameters such as particle size, concentration by measuring the temporal autocorrelation function[6,7]. The underlying purpose of this statistical approach (rather than the classical analysis of the relaxation dynamic [3]) is expected to enable coherent controlling and optical manipulation for realizing ultrafast relaxation dynamic, light extraction efficiency in some active optics like miniature laser, OLED and many more. Here, we analyze the relaxation process of a NLC in term of the DWS, and then develop TR-DWS to enable ultrafast linear E-O relaxation dynamic in a single LC cell at room temperature by controlling the orientational ordering of a NLC.

- ANALYTICAL METHOD

In a process of analyzing the relaxation dynamic of a NLC in terms of the DWS, it is worth mentioning that the temporal autocorrelation function is the only measurable quantity in DWS. In this section, we develop temporal autocorrelation function theoretically for a NLC system during its E-O relaxation process. Consider a plane polarized optical beam, $u(z,t) = A(z,\tau)\exp[i2\pi\nu t]$ with an optical frequency $\nu$ and field amplitude $A(z,\tau)$ passing through a NLC cell (planar), which is under the linear E-O relaxation process. It is a matter of fact that the optical beam gets randomized after propagating the field correlation length, which is $\xi \ (= \frac{d}{\pi})$ as stated above. If the total diffusing path length is $s$, the optical beam, during the relaxation process encounters total $N = s/\xi$ number of statistically independent dynamic phasors, each of which having a value $P(z_k) = (\sqrt{N})^{-1}|a_k|e^{i\vartheta_k}, k = 1,2,...N$ (**Fig.1b**), $|a_k|$ is the amplitude of k-th phasor and $\vartheta_k = \vartheta_m sin\left(\frac{z_k}{\xi}\right)$ is the phase at $z_k \in d$ (**Fig. 1c**)[1-3]. The transmitted intensity from the cell decays over time due to the time-evolution scattering of multiple number of random dynamic phasors $P(z_k)$. The field-autocorrelation function for this random dynamic process can be written in term of synergetic DWS [6-8] as:

$$g_1(\Gamma) = \exp[-2(\Gamma/\tau_0)N], \quad \quad \ldots (1)$$

and the intensity-autocorrelation function is expressed as [8],

$$g_2(\Gamma) = 1 + [g_1(\Gamma)]^2 \quad \quad \ldots (2)$$

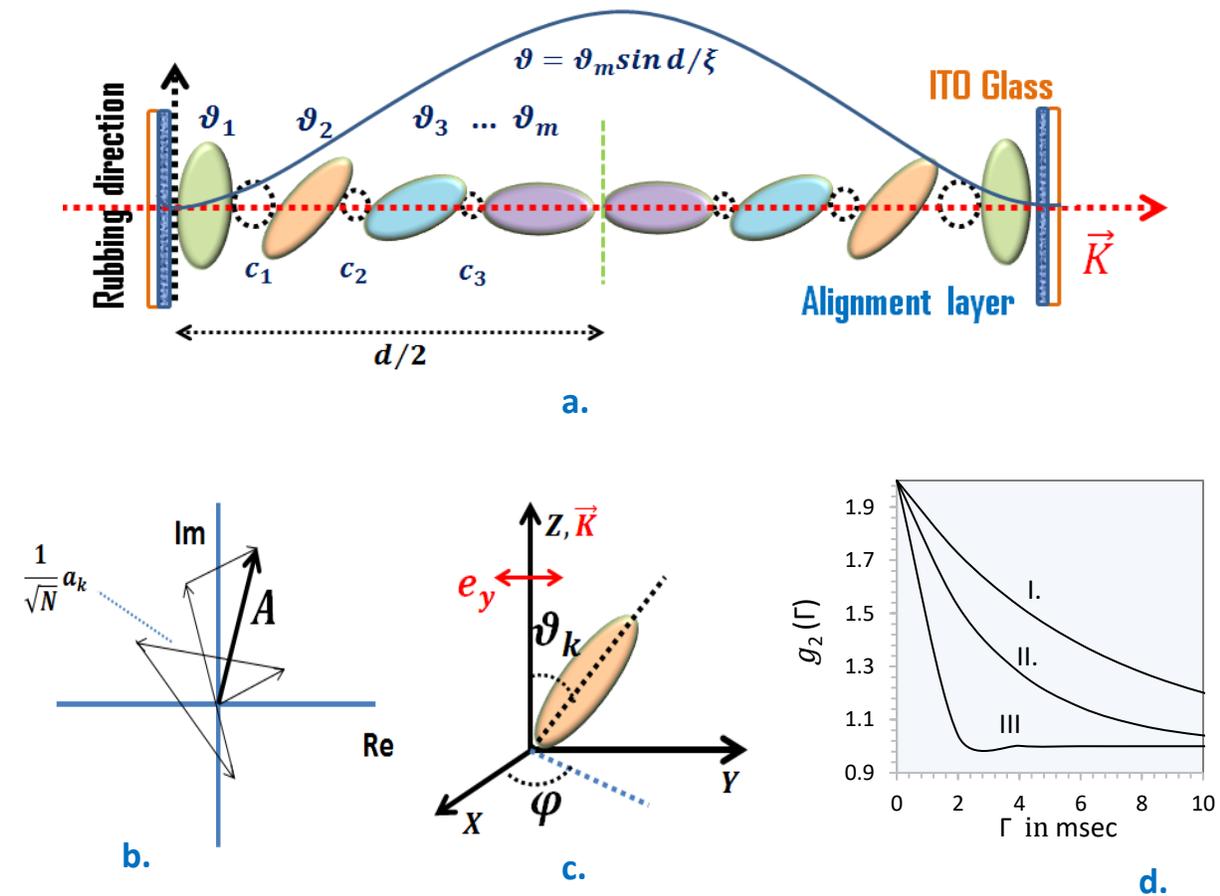

Fig.1a. The quasi-homeotropic state of a planar NLC system, comprised with liquid crystal molecule having $\Delta\varepsilon > 0$, in presence of an external electric field. The local directors are illustrated in various colors; $\vec{K}$ is the direction of propagation of optical beam. $C_n$ are the scattering centers arose due to the inhomogeneous distribution of surface energy along the cell thickness, b. random walk in the complex plane, c. polar and azimuthal angle of the k-th phasor in the cell geometry, d. a theoretical plot of Eq. 4, considering the characteristic relaxation time of a local director $\tau_0 = 1\mu sec$, and the number of scatterers for I. $N = 400\#$ II. $N = 800\#$, and III. $N = 1600\#$.

Analytical plots of Eq.2 are shown in Fig.1d for various values of $N$. It is important to validate these analytical plots by varying the number $N$ experimentally. Note that a set ($i$-th) of independent dynamic phasors $N_i$ is generated uncontrollably in a single NLC cell when the externally applied electric field transits from ON to OFF. But, generation of a multiple number of sets ($i = 1, 2, ..., n$) of independent dynamic phasors $N_{i(=1,2,...,n)}$ in a control way (to examine the Eq.2) may require a special design. In the following section, we design a cascaded-NLC system, comprised of multiple NLC cells, stacking one on top of other, and experimentally show the necessary condition to make the $N_{i(=1,2,...,n)}$ mutually independent. At the end of the section, we abbreviated the required electrical coding to control the generation of $N_{i(=1,2,...,n)}$ in real time.

- ## RESULTS AND DISCUSSION

Experimentally, it is observed that in the case of a stack comprised with two or even more number of NLC cells each of which having an identical orientation $\varphi$ (attributed by mechanical rubbing during the cell design, see supplementary information) in their directors $\hat{n}$ (Fig.2a-A1), the resultant relaxation time $\Gamma_R$ remained the same as that of a single NLC cell, which is typically $\Gamma_1 = 25$msec in our case. But, the same stack shows a faster $\Gamma_R$, if there is an azimuthal shift in the orientation of the cell directors by an amount $\Delta\varphi = \varphi' - \varphi$ from cell to cell (instead of letting them identical Fig.2a-A2). Experimentally, it is also observed that, $\Delta\varphi$ possesses a minimum value, which is $\Delta\varphi = 32° \pm 2°$, in order to observe a significant change in $\Gamma_R$ (Fig. 2b).

Recalling Eq.1, a possible statistical reason of the above $\varphi$ dependent $\Gamma_R$ is likely to be reasonable emphasizing the fact that, the minimum shift in $\Delta\varphi \ (= 32° \pm 2°)$ between any two neighboring cells in the cell stack (of $n$ number of cells) must give rise to a multiple sets ($i = 1,2,...,n$) of mutually independent dynamic (E-O relaxation dynamic) phasors $N_{i(=1,2...,n)}$. A microscopic route of this value $\Delta\varphi$ is further investigated experimentally by measuring the average span of the LC molecular distribution (see supporting information) inside the cell geometry. Interestingly, it is observed that the average span of molecular distribution estimates $\langle\varphi\rangle \cong 32°$ (Fig.2c). Because of that, stacking any two cells with a mutual azimuthal-offset $\Delta\varphi < \langle\varphi\rangle$ leads to an overlapping situation between the individual molecular distributions inside the two cells. This overlapping situation is equivalent to a single cell ($n = 1$), and effectively does not enable any additional independent dynamic phasors (apart from $N_{i=1}$) along the optical beam, hence the resultant relaxation time $\Gamma_R = \Gamma_1 \sim 25\, msec$ although, physically there are two cells. In contrast, stacking any two cells with a mutual azimuthal-offset $\Delta\varphi > \langle\varphi\rangle$, the resultant relaxation time becomes faster (i.e., $\Gamma_R = \Gamma_2 < \Gamma_1$) due to enabling two mutually independent sets of dynamic phasors $N_{i=1,2}$ along the optical beam. The maximum value of $n$ (the cell number) possible to stack in order to create the maximum sets of independent dynamic phasors can be estimated as $n = \varphi_T/\Delta\varphi = 6$, considering the value of the mutual azimuthal-offset $\Delta\varphi = 30°$, and the total span of the azimuthal angle $\varphi_T = 180°$ (instead of 360°) due to the symmetry of the nematic director $\hat{n} = -\hat{n}$ [1-3]. Based on the above understanding we finally design a cascaded-NLC system consists of 6 cells and the value of $\Delta\varphi$ kept to be 30° (Fig.2a-A3). The electrical coding of the cascaded-NLC is described in the following which further facilitates some real time applications.

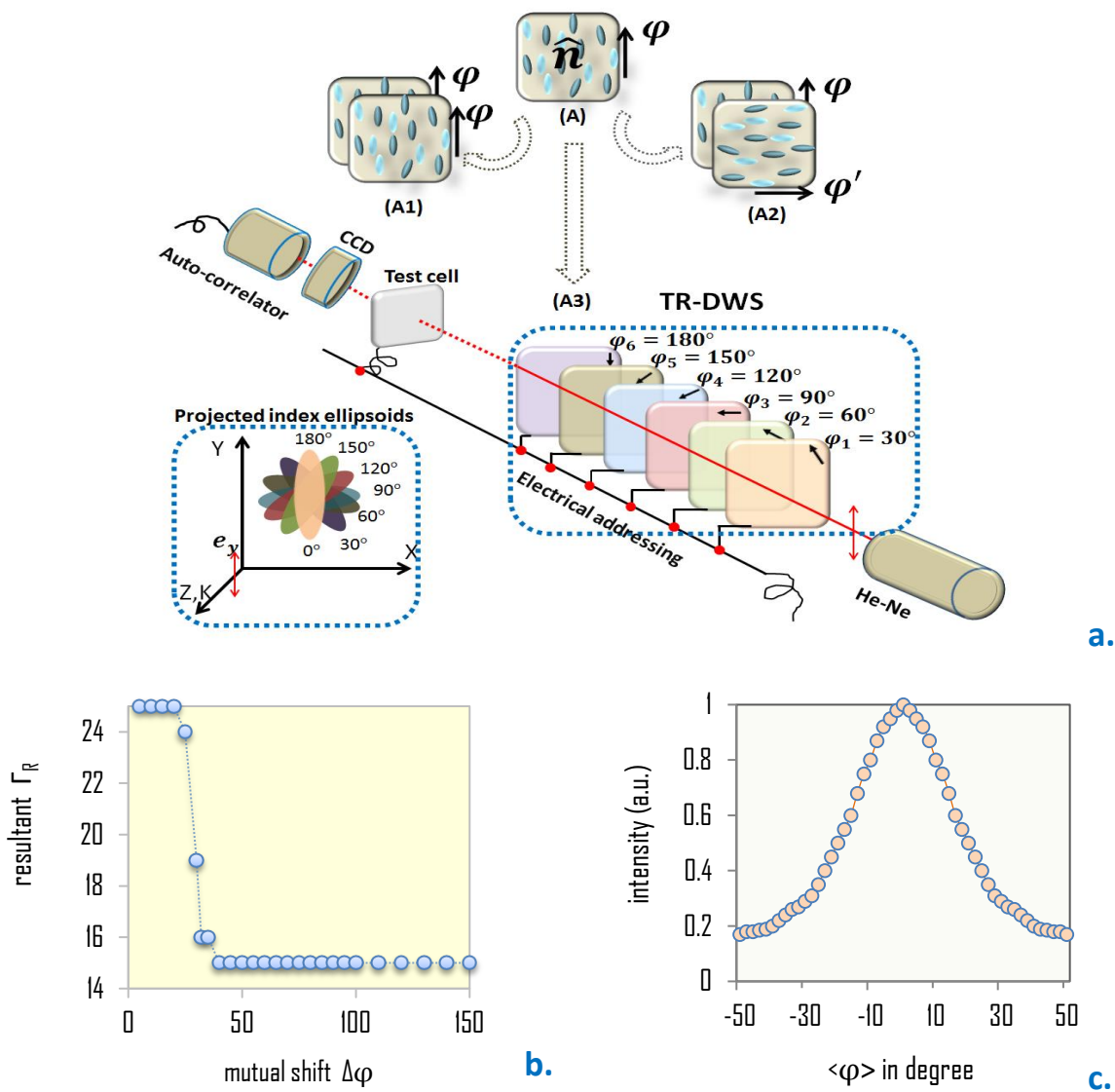

Fig.2a. illustrates (A) how LC molecules are aligned along the rubbing direction in a cell, (A1) two cells having same director orientations, (A2) two cells having a mutually shift in director orientations, (A3) cascaded-NLC system: TR-DWS, rubbing directions, indicated by arrow sign, are shifted azimuthally, the present electrical addressing shown is 111111, Inset: the X-Y projection of index ellipsoid of the TR-DWS. b. minimum offset in $\Delta\varphi$ required to have a significant temporal shift in a two-cell case, c. showing the variation in output intensity from the polarizing microscope in presence of a NLC cell over the rotating stage under continuous rotation; the transmitted intensity falls to 90% around 32° rotation of the stage, gives the measurement of the molecular distribution in a NLC.

Electrically each cell can be addressed either 1 or 0. A representative electrical configuration for the stack of 6 cells may be written as 000001, which stands for only a single cell relaxes (electric field transits from ON to OFF in it) out of 6 cells, hence only a single set ($i = 1$) of independent dynamic phasors $N_{i=1}$ is allowed to involve in the relaxation process. This essentially gives rise to the resultant relaxation time of the cascaded-NLC $\Gamma_R = \Gamma_1$ (against the coding 000001). In short, 000001≡ $\Gamma_1$, 000011 ≡ $\Gamma_2$... 111111 ≡ $\Gamma_6$, with $\Gamma_6 < \Gamma_5 < \cdots < \Gamma_2 < \Gamma_1$. Fig.3a shows how the resultant relaxation time $\Gamma_{R(=1,2,..,6)}$ changes by varying the electrical addressing or the effective number of cell in the cascaded-NLC system in real time. It is observed that the experimental relaxation curves in Fig.3b decay in a very similar way as that of the analytical plots in Fig.1d. This eventually confirms that the DWS is a suitable analytical probe to analyze the linear E-O relaxation dynamic in a NLC cell. Further, the ability to change the effective number of independent dynamic phasors $N_{i(=1,2,…,6)}$ in real time (through addressing the cells electrically), hence the temporal shift in $g_2(\Gamma)$ deciphering the fact that the cascaded-NLC can be used for spectroscopy, the TR-DWS. The working principle of TR-DWS is described below through an exemplary application.

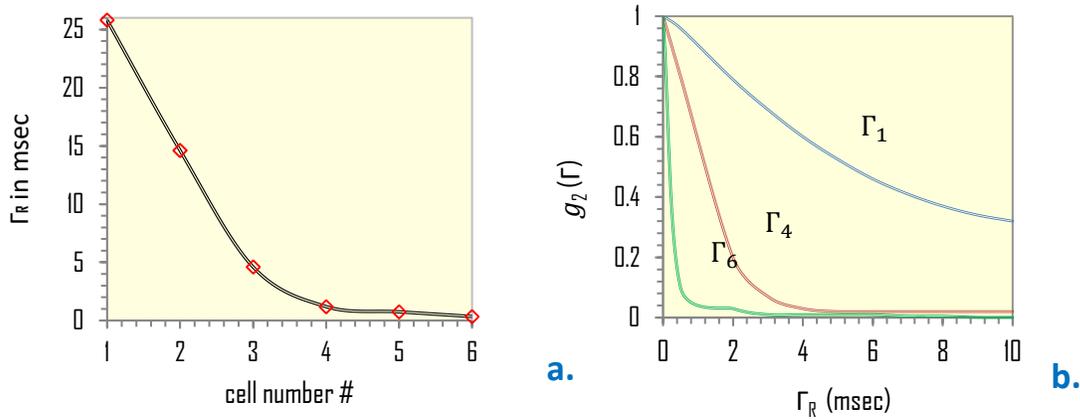

Fig.3a. showing the reultant relaxatin time getting faster with involving more number of cells to relax or in other word introducing more number of independent dynamic phasors along the optical beam through electrical coding, b. nature of the experimental decay curve with increasing the number of cell to relax in the cascaded-NLC system.

- APPLICATION

Ultrafast (ON and OFF field) dynamic in LCs is very demanding for various commercial applications [9-13]. But, mostly the relaxation dynamic of a classical NLC is limited to a few tens of millisecond [11-13]. The statistical reason is the long-range orientational correlation length of liquid crystals [4,5]. It is a matter of fact that chiral dopant induces a short-range (local) director orientation $\varphi$ in NLC [13,14]. Here, we use some chiral dopant to break the long-range correlation length in order to achieve an ultrafast dynamic in NLC. The application of the TR-DWS for this purpose is extremely significant and becomes apparent in the following section.

We begin to mix a small amount of chiral dopant in a NLC host and start monitoring the induced short range director orientation or, built-in $\varphi$ in the test NLC cell by the TR-DWS. Experimentally, the mixture of E7: R-811 ($\sim$ 99.7: 0.3 wt%) is filled in a non-rubbing cell of thickness $4\mu m$, and let it execute its relaxation dynamic. Exploiting Eq.1 one can estimate the dynamic phasors $N$, involved in the relaxation process (of the composite E7: R-811) with respect to the reference temporal signal $\Gamma_{R(=1,2,…,6)}$, generated from the well-defined cascaded-NLC system. In order to do so, we generate a set of characteristic $\Gamma_{R(=1,2,…,6)}$ iteratively from the cascaded-NLC using electrical addressing, starting from $000001 \equiv \Gamma_1$, and probe onto the test cell (E7: R811). At each iterative step, our main focus is to notice if there is any temporal shift introduced in the reference $\Gamma_{R(=1,2,…,6)}$. A temporal shift in

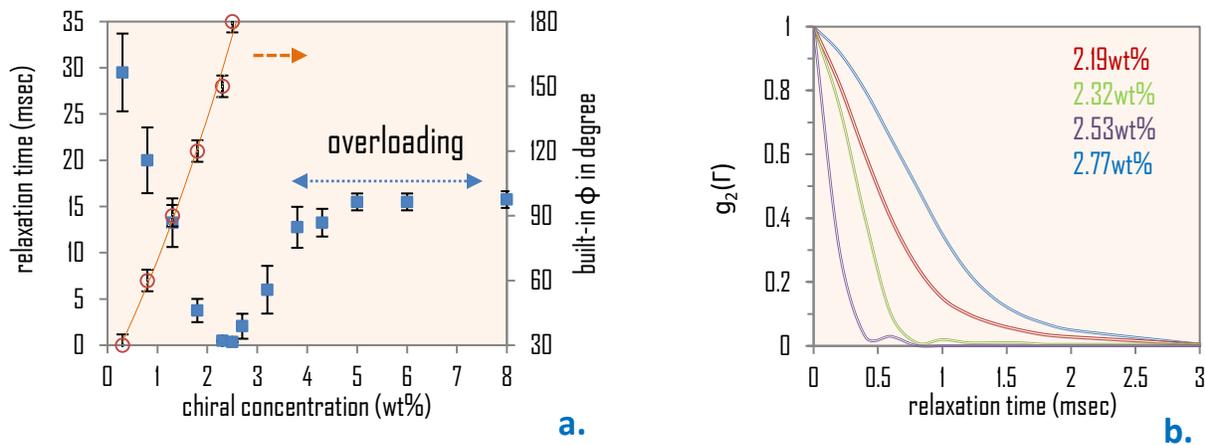

a.

b.

Fig.4a. Control curve of the relaxation mechanism by optimizing the chiral dopant concentration with the help of TR-DWS, to estimate the relative built in $\varphi$ in test cell, and overloading effect of chiral dopant, b. Ultrafast linear E-O relaxation time in a single liquid crystal cell, different curves corresponding to various wt% of chiral dopant.

reference $\Gamma_{R(=1,2,…,6)}$ definitely signifies the presence of some additional (apart from the dynamic phasors $N_{i(=1,2,…,6)}$ involved in the cascaded-NLC) independent dynamic phasors in the test cell as in Eq.1. No shift in $\Gamma_{R(=1,2,…,6)}$ signifies the fact that the set of dynamic phasors involved in the reference system (cascaded-NLC) at that particular iterative step is identical to that of the test cell or in other word, the built-in $\varphi$ in the test cell at that circumstance is exactly same as that of the total orientational diversity possessed in the reference cascaded-NLC which is $\Delta\varphi *$(the number of cells executing the relaxation dynamic at that iterative step). After a first iterative run (from 000001 through 111111), it is possible to identify the span of the built-in $\varphi$ in the test cell (see supplementary information). Depending on the finding of the built-in $\varphi$ from the 1st run of the complete iterative process, further concentration of the chiral dopant is decided towards a target to incorporate the total built-in $\varphi_T = 180°$ (Fig.4a), to see if we have the fastest relaxation time in the test cell. It is observed that at 2.53wt% of chiral dopant, the relaxation time gets improved from 25msec to 350$\mu sec$(Fig.4b). It is also observed that further increasing the chiral concentration to 2.77wt% results in a slower relaxation dynamic (Fig.4a). This may be due to some overloading phenomenon. The typical region corresponding to the fastest relaxation time in Fig.4a is too narrow and sensitive to control the chiral concentration without visualizing them progressively with our TR-DWS. Hence, the underlying purpose of using the TR-DWS is to enable the indispensable control of visualizing the span of the built-in $\varphi$ due to the chiral dopant. It is noted that the conventional polarizing microscope cannot visualize the built-in $\varphi$ (through measuring the change in birefringence), as the cell is observed to be in highly scattering state at this low concentration regime of chiral dopant.

- # CONCLUSION

In conclusion, the relaxation dynamic of NLC is analyzed in terms of DWS. Analytically, it is found that the temporal-autocorrelation function has a dependency on the number of independent dynamic phasors involved in the relaxation process. Experimentally, we validate the same through designing a multiple cell stack to achieve a control over the number variation of the independent dynamic phasors thus, does the temporal-autocorrelation function. The prime advantage of this statistical approach of understanding the relaxation dynamic of NLC helps us visualizing the progressive change in orientational ordering in a test NLC cell upon mixing some chiral dopant purposefully; to shorten the long orientational range, hence improve the response time. Finally, a high speed window in a LC at room temperature has been possible to open up through our advanced spectroscopic technique, named TR-DWS.

**Acknowledgement**
We would like to acknowledge many discussions with Prof. Herve Folliot. This research was supported by TVSF collaborative research grant.

**Correspondence** [*]skmanna@ucdavis.edu